
\magnification=1200
\catcode `@=11
\hsize 16truecm
\vsize 24truecm
\def\build#1_#2^#3{\mathrel{\mathop{\kern 0pt#1}\limits_{#2}^{#3}}}

\font\eightrm=cmr8

\font\twelve=cmbx10 at 13pt
\font\ten=cmbx12

\def\n{\noindent}
\def\s{\smallskip}
\def\m{\medskip}
\def\b{\bigskip}

\def\tv{\tvi\vrule}
\font\ten=cmbx10 at 13pt
\font\twelve=cmbx10
\font\eight=cmr8
\baselineskip 15pt

{
\centerline{\twelve Centre de Physique Th\'eorique - CNRS - Luminy,
Case 907}
\centerline{\twelve F--13288 Marseille Cedex 9 - France }
\centerline{\bf Unit\'e Propre de Recherche 7061}

\vskip 2truecm

\centerline{\ten  TEXTURE SEGMENTATION BY LOCAL}
\centerline{\ten BI-ORTHOGONAL DECOMPOSITION}
\bigskip
\centerline{{\bf J.A. DENTE}\footnote{$^{\star}$}{\eight
Laborat\'orio de Mecatr\'onica Dept. Eng. Electrot\'ecnica, Instituto
Superior T\'ecnico,}\footnote{}{\eightrm av. Rovisco Pais 1, 1096
Lisboa Codex Portugal}{\bf,  R.
LIMA}\footnote{$^{\dagger}$}{\eight Centre de Physique Th\'eorique, CNRS -
Luminy, Case
907, 13288 Marseille Cedex 9}{\bf and R. VILELA MENDES$^{\star}$}}

\vskip 2truecm

\centerline{\bf Abstract}

\medskip

We investigate the ability of a local bi-orthogonal decomposition to
build texture segmentation of images. Using the structures
associated to the local decomposition of the image independent row
and columns we perform a segmentation, where the regions are defined
by the property of having a smooth variation of the corresponding
entropy. Examples are choosen in texture made and also in real life
images. The size of the local analysis is also determined by the
properties of the (global) bi-orthogonal decomposition.

\vskip 3truecm


\bigskip

\noindent Number of figures : 18

\bigskip

\noindent January 1994

\noindent CPT-94/P.3010

\bigskip

\noindent anonymous ftp or gopher : cpt.univ-mrs.fr

\footline={}

\vfill\eject    }

\pageno=1

\item{\hbox to\parindent{\enskip {\bf 1.} \hfill}} {\bf TEXTURE
SEGMENTATION AND THE LOCAL BI-ORTHOGO\-NAL DECOMPOSITION}

\vglue 0,2truecm

{\it Image segmentation} is the partition of a plane image into
exclusive regions which, in some sense, are homogeneous. When the
purpose of segmentation is to distinguish some object from a
background and the brightness of the object and the background are
significantly different, segmentation by {\it gray level
thresholding} is a possibility. This works well for example in
automated manufacturing processes if the assembly parts are kept dark
against a bright  background but, in most images of 3-dimensional
objects, different illumination levels in different parts of the same
object make gray level thresholding a poor discriminating technique.

\s

A more frequently used technique is {\it edge detection} by
convolution of the image with a discrete difference operator,
followed by a contour filling algorithm to decide which pixels belong
to each one of the segmented regions. Edge detection also faces
serious difficulties because factors such as illumination may either
hide physical boundaries or, through shadows, cause brightness
discontinuities which are not related to any real boundaries.

\s

The reason why segmentation in computer vision is such a difficult
problem, as compared with the ease with which the ``eye plus brain''
system performs this task, is because in the brain an huge amount of
information is stored concerning the way the real world looks like.
Based on a few external stimuli, like brightness levels and a few
contours, the bulk of the segmentation process in the brain is likely
to be mostly an exercise in pattern matching of the external stimuli
with our ``image of the world'' data basis. While our computers
are not equipped with a data basis of comparable size and
complexity as the brain, computers must rely on a refinement in the
analysis of the external stimuli part of the process. This means that
quantitative characterizations of global and local properties of the
image must be developed, which might even have to be finer and more
accurate than those performed by the human eye. Only then, might we
compensate for the weakness of the data basis in computer vision.

\s

The most difficult of all segmentation problems occurs when
different regions of the image cannot be distinguished by gray level
nor by sharp boundaries, but only by a difference in texture. Texture
refers to the {\it local} characteristics of the image. A local gray
level histogram is a local statistical parameter. However, for the
texture, what matters most are the local spatial correlations between
pixel intensities. The brain will probably perform the {\it texture
segmentation} task by pattern matching with its data basis but, in the
computer, the only alternative is to attempt an objective
mathematical  characterization of what texture means. Several
quantities have been proposed as a measure of texture. For example~:

\m

\item{\hbox to\parindent{\enskip --- \hfill}} {\it The gray level
co-occurrence matrix} $P^{[1,2]}$ is a matrix with elements $P(i,j)$
which are the number of pairs of pixels that in some neighborhood
have intensities $i$ and~$j$.

\item{\hbox to\parindent{\enskip --- \hfill}} {\it The local
autocorrelation function$^{[3]}$}
$$A_I(m,n)=\sum_j\sum_k I(j,k)I(j-m,k-n)$$

\n where $I(i,j)$ is the image intensity at the point with coordinates
$(i,j)$ and the sum is over a small window $-D\le m,n\ge D$, or
quantities derived from $A_I(m,n)$, like the {\it autocorrelation
spread measures$^{[4]}$}.

\item{\hbox to\parindent{\enskip --- \hfill}} The {\it number of
edges}$^{[3]}$ in a neighborhood

\item{\hbox to\parindent{\enskip --- \hfill}} The {\it local Fourier
spectrum$^{[5,6]}$}

\item{\hbox to\parindent{\enskip --- \hfill}} The {\it singular value
decomposition}$^{[7]}$ of local texture samples

\item{\hbox to\parindent{\enskip --- \hfill}} The {\it moments of the
gray-level histograms} of small windows

\item{\hbox to\parindent{\enskip --- \hfill}} {\it Texture
primitives} and {\it grammar rules} to generate the pattern

\s

In this paper we are proposing and testing the idea that the
entropy associated to the bi-orthogonal decomposition is an adequate
parameter to characterize different textures. The bi-orthogonal
decomposition (see Appendix A and Ref.~[13] for more details) is a
2-dimensional generalization of the Karhunen-Lo\`eve$^{[8,9]}$
technique which states that a real signal u(x,y) on two variables may
be uniquelly decomposed in the form
$$u(x,y)=\sum_k\alpha_k\phi_k(x)\Psi_k(y)\eqno(1)$$
where $\{\phi_k(x)\}$ and $\{\Psi_k(y)\}$ are both orthonormal sets.
The expansion basis is generated by the signal itself and the pairs
$\phi_k(x)\Psi_k(y)$ are the independent $x,y$-structures that
compose the image. They encode the full nature of the
geometrical 2-dimensional correlations in the image. Two-dimensional
spatial  correlations between pixel intensities being at the very
root of the notion of texture, it is natural to conjecture that the
bi-orthogonal decomposition of local blocks of appropriate size is an
appropriate tool to characterize textures in an image. We therefore
propose the following three-step process for texture segmentation by
local bi-orthogonal decomposition (LBOD)~:

\b

\parindent=22pt
\item{\hbox to\parindent{\enskip {\bf i)} \hfill}} {\bf Identify
the texture average scale}

\vglue 0,2truecm

Compute the Fourier transform of a few randomly chosen lines
and columns of the image. Next the algorithm should identify the first
peak in the spectrum after the peak around zero (which corresponds to
the average pixel intensity and long-range slow variations). In
typical images the first large peak away from zero is the lowest
texture frequency $\omega_T$. A block size $M\times N$ is then chosen
where $M$ and $N$ correspond to the average $1/\omega_T$ along the
lines and the  columns. Instead of using the Fourier transform of a
set of lines and columns, we may use the Fourier transform of one of
the modes in a global bi-orthogonal decomposition of the image.

\b

\item{\hbox to\parindent{\enskip {\bf ii)} \hfill}} {\bf Construct the
entropy image}

\vglue 0,2truecm

The image is now divided into blocks of size $M\times N$ and the
bi-orthogonal entropy (Eq. B.2) of each block is computed. (For a
comparison of the bi-orthogonal entropy with other entropy notions
see Appendix B). Assigning to each block its entropy value one
obtains a {\it block entropy image}.

\b

\item{\hbox to\parindent{\enskip {\bf iii)} \hfill}} {\bf Segmentation from
the entropy image}

\vglue 0,2truecm

The entropy image is smoothed by some standard algorithm and
contour tracing from the smoothed entropy image completes the process
of texture segmentation by LBOD.

\b

\parindent=20pt

\item{\hbox to\parindent{\enskip {\bf 2.} \hfill}} {\bf EXAMPLES}

\vglue 0,2truecm

Before dealing with real world images we tested the algorithm on
the image shown in figure 1. This image has several textures which
were constructed in such a way that the local (in $8\times 8$ blocks)
gray level average is everywhere the same and no boundary lines exist
separating the different textures. In this sense this example
presents a pure case of segmentation by textures.

\s

The average texture scale was found by computing the Fourier
transform of the eigenfunctions $\phi_k$ of the global
bi-orthogonal decomposition. Figure 2 shows the spectrum of
$\phi_2$. A large peak may be seen, that corresponds to a block of
dimension $8\times 8$ pixels.

\s

The image is then divided into blocks of size $8\times 8$ and the
bi-orthogonal entropy of each block is computed to obtain the block
entropy image shown in figure 3.

\s

In the entropy image, zones with different textures are well
separated by the entropy values. This enables us to use a simple
gradient algorithm to find the contours, thus performing the texture
segmentation. Figure 4 shows the result of this operation.

\s

To obtain an entropy image with better resolution we might compute
the entropy in a neighbourhood of every pixel in the original image.
However this procedure is time consuming. It suffices to generate an
entropy image using the block entropy for only a smaller number of
pixels in the original image. Figures 5a,b show the entropy image
evaluated using $8\times 8$ blocks separated by 4 pixels. Figure 6
shows the contours obtained in this case.

\s

If the local entropy actually characterises the local texture, it
should not be too sensitive to illumination levels in the image. We
have tested this feature by  changing the intensity in one half of
our test image (Figure 7).

\s

The relative insensitivity of the block entropies to illumination
levels is apparent from the entropy and contour images shown in the
figures 8 and 9.

\s

In the ``pure textures'' example described above,
texture segmentation by local bi-orthogonal decomposition seems to
work efficiently. In real world images, however, we see some
difficulties and limitations of the method. Take for example the
``bears'' image of figure 10.

\s

The first difficulty occurs in the choice of the block size. In a
real world image  many different texture scales may occur, hence
there is no unique block size appropriate for all texture features.
Figure 11 shows the spectrum of the global $\phi_2$ eigenfunction of
the ``bears'' image. A large dimension is suggested for the dominant
block size. This correspond to large areas of the image (the water
and  the mountain) which are not the main objects in the image. If
other eigenfunctions are used other characteristic block sizes are
found, as shown in the figures 12 and 13. The low block sizes are
associated to low energy levels, but in spite of this their small
intensities they are important to define the details of the image,
that is the micro-structures that our brain understands.

\s

The effect of the block size is shown in the entropy images of
figures 14 and 15,which use blocks of size ($18\times 18$) and
($4\times 4$) respectively.

\s

Contours are difficult to obtain in a simple way, because the
entropy image has many different values as shown in three dimension
figures 16 and 17. With a simple gradient plus clipping algorithm one
obtains the result show in figure 18.

\b

\n{\bf Appendix A~:} {\it The bi-orthogonal decomposition}

\vglue 0,2truecm

The decomposition into orthogonal modes, of probability theory, is
a well known procedure in signal analysis referred to as
Karhunen-Lo\`eve decomposition$^{[8,9]}$ or principal component
analysis. Given a random vector $x_i\{i=1,\dots,N\},\ x_i\in X$, the
covariance matrix $Q=\left[xx^T\right]$ is diagonalized and the random
vector $x$ expressed as
$$x=\sum_{i=1}^N \alpha_i\phi_i\eqno(\hbox{A}.1)$$
where $\phi_i$ are the eigenvectors of $Q$, i.e. the columns of the
matrix $A$ that diagonalizes $Q(Q=A\lambda A^T,\ \lambda$ diagonal).
The best (mean-square) $P$-component approximation to the signal $x$
(with $P<N$) is obtained choosing the $\alpha_i$ coefficients
associated with the largest $P$ eigenvalues. This property makes the
Karhunen-Lo\`eve decomposition a standard data compression technique.
The  Karhunen-Lo\`eve technique has been used for image
processing$^{[10-12]}$. The  image is divided into small blocks, each
block is treated as a sample of an one-dimensional statistical
signal, the labeling of the blocks playing the role of time
variable. The expectation value in the covariance
$E\left[xx^T\right]$ is then taken over these blocks.

\s

However, in a image, an important part of the relevant
informationis related to  geometrical correlations. They concern the
variation of the gray levels along particular directions and define
the contour and the shape information content of the image. This
suggests the use, for image processing, of a generalization of the
Karhunen-Lo\`eve technique where bidimensional correlations are
explicitly taken into account.

\s

The bi-orthogonal decomposition analyses signals $u(x,y)$ that
depend on  variables defined in two different spaces $(x\in X,\
y\in Y)$. We summarize below  the main results concerning the
bi-orthogonal decomposition and refer to [13] for more details.

\s

Let the signal $u(x,y)$ be a measurable complex-value function
defined on $X\times Y$, where $X$ and $Y$ are either $R^n$ or $Z^n$ or
subsets of one of these. The signal defines a linear operator $U$~:
$L^2(Y)\to L^2(X)$ by
$$(U\Psi)(x)=\int_y u(x,y)\Psi(y)dy\quad\forall\Psi\in
L^2(Y)\eqno(\hbox{A.2a})$$
with adjoint operator $U^{\dag}$~: $L^2(X)\to L^2(Y)$
$$(U\phi)(y)=\int_y u^{\star}(x,y)\phi(x)dx\quad\forall\phi\in
L^2(X)\eqno(\hbox{A.2b})$$

The analysis of the signal $u(x,y)$ is the spectral analysis of the
operator $U$. In general the spectrum contains continuous and point
spectral components. However we will assume that $u\in L^2(X\times
Y)$ or that $X$ and $Y$ are compact and $u$ continuous implying that
$U$ is a compact operator. Then the spectrum consists of a countable
set of isolated points. There is a canonical decomposition of
$u(x,y)$ such that
$$u(x,y)=\sum_{k=1}^{\infty}\alpha_k\phi_k(x)\Psi_k^{\star}(y)
\eqno(\hbox{A}.3)$$
is norm-convergent
$$\alpha_1\ge\alpha_2\ge\dots\ge\alpha_k\ge\dots >0,\ \lim\alpha_k=0,
\quad(\phi_i,\phi_j)=(\Psi_i,\Psi_j)=\delta_{i,j}$$

The functions $\phi_k(x)$ are also eigenfunctions of the operator
$L=UU^{\dag}$, and the $\Psi_k(y)$ are eigenfunctions of
$R=U^{\dag}U$. These functions are related by the following equation
$$\phi_k=\alpha_k^{-1}U\Psi_k\eqno(\hbox{A}.4)$$

The operators $L$ and $R$ are non-negative operators with kernels
$l(x_1,x_2)$ and $r(y_1,y_2)$ that are the $X$ and $Y$-correlation
functions of the signal
$$l(x_1,x_2)=\int_y u(x_1,y)u^{\star}(x_2,y)dy\eqno(\hbox{A}.5)$$
$$r(y_1,y_2)=\int_x u^{\star}(x,y_1)u(x,y_2)dx\eqno(\hbox{A}.6)$$

The eigenvectors $\phi_k$ and $\Psi_k$ of the $L$ and $R$ operators
appear, in the  decomposition (A.3) of the signal, intrinsically
coupled to the same eigenvalue $\alpha_k^2$. The
products $\phi_k\Psi_k$ are therefore the independent $X,Y$ structures
that compose the signal. This decoupling of structures occurs
because, as opposed to other methods of signal analysis (Fourier,
wavelets, etc), the functional basis decomposing $u$ is produced by
$u$ itself.

\s

{}From the bi-orthogonal decomposition one may construct
several global quantities~:

\s

\n The square of the norm of the signal in
$L^2(X\times Y)$, which we call the energy, equals the sum of the
eigenvalues $$E(u)=\int_{X\times
Y}u(x,y)u^{\star}(x,y)dxdy=\sum_k\alpha_k^2\eqno(\hbox{A}.7)$$
Similarly one defines $X$-dependent and $Y$-dependent energies as
$$E_x(u)=\int_Y u(x,y)u^{\star}(x,y)dy=\sum_k
\alpha_k^2|\phi_k(x)|^2\eqno(\hbox{A}.8)$$
$$E_y(u)=\int_X u(x,y)u^{\star}(x,y)dx=\sum_k
\alpha_k^2|\Psi_k(y)|^2\eqno(\hbox{A}.9)$$

The dimension of a signal is defined to be the dimension of the
range of $U$. For the compact case this is the number of non zero
eigenvalues $\alpha_k^2$.

\s

The $\varepsilon$-dimension of the signal is the number
of eigenvalues larger than $\varepsilon$. The size of the eigenvalues
is a good characterization of the degree of approximation
in the sense that, truncating the $U$ operator to
$$U_p=\sum_k^p\alpha_k\phi_k\Psi_k$$
the norm of the error $||U-U_p||$ is smaller than the first neglected
eigenvalue. The notion of $\varepsilon$-dimension is useful to
characterize noisy signals.

\b

\n{\bf Appendix B~:} {\it Entropy}

\vglue 0,2truecm

The notion of entropy may be used to estimate the information
content. It measures the amount of disorder in a system and, in this
sense, it is sensitive to the spread of possible states which a
system can adopt. For an image the simplest idea is to make these
states correspond to the possible values which individual pixels can
adopt. Then, the entropy (associated to the gray level histogram)
would be given by
$$E=-\sum_{j=0}^{M-1}P(j)\log P(j)\eqno(\hbox{B}.1)$$
where $P(j)$ is the probability of pixel value $j$ and $M$ is the
number of different values which the pixels can take. Equation
(B1) represents the information  content of the image only if all
pixels are uncorrelated. This is not the case in real world images.
Consider, instead of the original image, an image formed by the
differences of neighbouring pixels. The original image can be
reconstructed from the difference image together with the value of
the first pixel. Therefore they contain the same information. However
one usually finds that the entropy of the gray level histogram of the
``difference image'' is smaller than the one for the original image.
This occurs because the difference image extracts some of the space
correlations existing in the image, hence its entropy is closer to
the  actual information content of the image.

\s

As explained in Appendix A, the bi-orthogonal decomposition extracts
the  normal modes of the image fully taking into account
the correlations along the  two coordinate axis. Therefore we expect
that the entropy associated to the weights of the modes in the
bi-orthogonal decomposition would be even closer to the actual
information content of the image.

\s

Associated to the eigenvalue structure of the bi-orthogonal
decomposition we define an entropy by
$$H(u)=-\lim_{N\to\infty}{1\over \log N}\sum_{k=1}^N p_k\log
p_k\eqno(\hbox{B}.2)$$
where
$$p_k={\alpha_k^2\over
\displaystyle\sum_k\alpha_k^2}\eqno(\hbox{B}.3)$$
and $X$ and $Y$-entropies by
$$H_x(u)=-\lim_{N\to\infty}{1\over \log
N}\sum_{k=1}^N p_k(x)\log p_k(x)\eqno(\hbox{B}.4)$$
$$H_y(u)=-\lim_{N\to\infty}{1\over \log N}\sum_{k=1}^N p_k(y)\log
p_k(y)\eqno(\hbox{B}.5)$$
where
$$p_k(x)={\alpha_k^2|\phi_k(x)|^2\over
\displaystyle\sum_k\alpha_k^2|\phi_k(x)|^2}\quad
;\quad p_k(y)={\alpha_k^2|\Psi_k(y)|^2\over
\displaystyle\sum_k\alpha_k^2|\Psi_k(y)|^2}$$

We have computed the gray level histogram entropy, the entropy of the
difference images and the entropy of the bi-orthogonal decomposition
for real  world images and for our textures test image. For
real world images we tipically  find that the entropy of the
difference image is smaller than the gray level histogram entropy. An
exception is our textures test image. This is because the local
textures lead to strong local fluctuations at the pixel level. In all
cases however the bi-orthogonal entropy is the smaller of them all.
In table I we list the computed values for the ``bears'' image and the
textures test image.

\b

\centerline{{\bf Table} - Entropy values for test and ``bears''
images.}

\vglue 0,2truecm

$$\vbox{
\def\tvi{\vrule height 12pt depth 5pt width 0pt}
\offinterlineskip\halign{
\tv\quad#\quad\hfill\tv& \quad\hfill#\quad\tv&
\quad\hfill#\quad\tv&
\quad\hfill#\quad\tv\cr
\noalign{\hrule}
\hfill{\it Image\/}& {\it Bi-orthogonal\/}\hfill& {\it
Histogram\/}\hfill& {\it Difference\/}\hfill\cr
\noalign{\hrule}
textures& 0.1765& 0.3572& 0.5009\cr
bears& 0.0780& 0.5642& 0.3208\cr
\noalign{\hrule}
}}$$

\vfill\eject

\n{\bf REFERENCES :}

\vglue 0,2truecm

\parindent 1truecm

\item{\hbox to\parindent{\enskip \hphantom{0}[1] \hfill}}
R.M.~Haralick and L.G.~Shapiro~; ``Glossary of computer vision
terms'', Pattern Recognition, {\bf 24} (1991) 69.

\item{\hbox to\parindent{\enskip \hphantom{0}[2] \hfill}}
R.M.~Haralick, K.~Shanmugan and I.~Dinstein~; ``Texture features for
image  classification'', IEEE Trans. Systems, Man and
Cybernetics, {\bf 3} (1973) 610.

\item{\hbox to\parindent{\enskip \hphantom{0}[3] \hfill}}
A.~Rosenfeld and E.B.~Troy~; ``Visual texture analysis'', Proc.
UMR-Mervin, J.~Kelly Communications Conference Section, {\bf 10-1},
1970.

\item{\hbox to\parindent{\enskip \hphantom{0}[4] \hfill}}
O.D.~Fangeras and W.K.~Pratt~; ``Decorrelation methods of texture
feature  extraction'', IEEE Trans. Pattern Analysis and Machine
Intelligence, {\bf 2} (1980) 323.

\item{\hbox to\parindent{\enskip \hphantom{0}[5] \hfill}}
G.G.~Lendaris and G.L.~Stanley~; ``Diffraction pattern sampling for
automatic pattern recognition'', Proc. IEEE, {\bf 58} (1970) 198.

\item{\hbox to\parindent{\enskip \hphantom{0}[6] \hfill}}
A.~Rosenfeld~; ``Automatic recognition of basic terrain types from
aerial photographs'', Photogrammic Engineering, {\bf 28} (1962) 115.

\item{\hbox to\parindent{\enskip \hphantom{0}[7] \hfill}}
B.~Ashjari~; ``Singular value decomposition texture measurements for
image  classification'', Thesis, Univ. Southern California, 1982.

\item{\hbox to\parindent{\enskip \hphantom{0}[8] \hfill}} M.~Lo\`eve,
``Probability theory'', Van Nostrand, 1955.

\item{\hbox to\parindent{\enskip \hphantom{0}[9] \hfill}}
K.~Karhunen, Ann. Acad. Sci. Fennicae, Ser. A1, Math. Phys., {\bf 37}
(1946) 1.

\item{\hbox to\parindent{\enskip [10] \hfill}} J.S.~Lim,
``Two-dimensional signal and image processing'', Prentice-Hall, 1988.

\item{\hbox to\parindent{\enskip [11] \hfill}} T.D.~Sanger, ``Optimal
unsupervised learning in feedforward neural networks'',
MIT Artificial Intelligence Laboratory Tech. Report 1086, 1989.

\item{\hbox to\parindent{\enskip [12] \hfill}} T.D.~Sanger, Neural
Networks, {\bf 2} (1989) 459.

\item{\hbox to\parindent{\enskip [13] \hfill}} N.~Aubry, R.~Guygnnet
and R.~Lima, J. Stat. Phys., {\bf 64} (1991) 683.

\item{\hbox to\parindent{\enskip [14] \hfill}} T.~Pun, Signal
Processing, {\bf 2} (1980) 223, {\bf 2} (1981) 210.

\item{\hbox to\parindent{\enskip [15] \hfill}} J.N.~Kapur, P.K.~Sahoo
and A.K.C.~Wong, Computer Graphics, Vision and Image Processing, {\bf
29} (1985) 273.

\vfill\eject

\n{\bf Figure Captions}

\vglue 0,2truecm

\parindent 2truecm

\item{\hbox to\parindent{\enskip Fig. 1 \hfill}} ($144\times 272$)
pixel image with uniform average gray level in $8\times 8$ blocks.

\item{\hbox to\parindent{\enskip Fig. 2 \hfill}} Spectrum of the
global $\phi_2$ eigenfunction.

\item{\hbox to\parindent{\enskip Fig. 3 \hfill}} Entropy image for
non-overlapping $8\times 8$ blocks

\item{\hbox to\parindent{\enskip Fig. 4 \hfill}} Contours of the
entropy image using non-overlapping blocks of size $8\times 8$.

\item{\hbox to\parindent{\enskip Fig. 5 \hfill}} Entropy image
using blocks of size $8\times 8$ separated by 4 pixels.

\item{\hbox to\parindent{\enskip Fig. 6 \hfill}} Contours of the
entropy image shown in the figure 5.

\item{\hbox to\parindent{\enskip Fig. 7 \hfill}} Test image with
two illumination levels.

\item{\hbox to\parindent{\enskip Fig. 8 \hfill}} Block entropy for
the test image of figure 7.

\item{\hbox to\parindent{\enskip Fig. 9 \hfill}} Contours obtained
{}from the entropy image in figure 8.

\item{\hbox to\parindent{\enskip Fig. 10 \hfill}} Bears. A real
world test image ($240\times 320$).

\item{\hbox to\parindent{\enskip Fig. 11 \hfill}} Spectrum of the
global $\phi_2$ eigenfunction in the ``bears'' image.

\item{\hbox to\parindent{\enskip Fig. 12 \hfill}} Spectrum of the
global $\phi_3$ eigenfunction in the ``bears'' image.

\item{\hbox to\parindent{\enskip Fig. 13 \hfill}} Spectrum of the
global $\phi_{60}$ eigenfunction in the ``bears'' image.

\item{\hbox to\parindent{\enskip Fig. 14 \hfill}} ``Bears'' entropy
image with blocks of size $18\times 18$ separated by 4 pixels.

\item{\hbox to\parindent{\enskip Fig. 15 \hfill}} ``Bears'' entropy
image with blocks of size $4\times 4$ separated by 1 pixel.

\item{\hbox to\parindent{\enskip Fig. 16 \hfill}} Three-dimensional
$18\times 18$ entropy image.

\item{\hbox to\parindent{\enskip Fig. 17 \hfill}} Three-dimensional
$4\times 4$ entropy image.

\item{\hbox to\parindent{\enskip Fig. 18 \hfill}} Contours obtained
{}from the $4\times 4$ entropy image.

\end